\documentclass[aps,preprint]{revtex4-1}
\usepackage{graphicx}
\usepackage{amsmath}
\usepackage{makeidx}
\usepackage{amsfonts}
\usepackage{amssymb}
\usepackage{tabularx}
\usepackage{float}

\begin{document}

\title{Quantum walk in a reinforced free-energy landscape: Quantum annealing with reinforcement} 

\author{A. Ramezanpour}
\email{aramezanpour@gmail.com}
\affiliation{Department of Physics, School of Science, Shiraz University, Shiraz 71946-84795, Iran}
\affiliation{Leiden Academic Centre for Drug Research, Faculty of Mathematics and Natural Sciences, Leiden University, PO Box 9500-2300 RA Leiden, The Netherlands}

\date{\today}

\date{\today}

\begin{abstract}
Providing an optimal path to a quantum annealing algorithm is key to finding good approximate solutions to computationally hard optimization problems. Reinforcement is one of the strategies that can be used to circumvent the exponentially small energy gaps of the system in the annealing process. Here a time-dependent reinforcement term is added to the Hamiltonian in order to give lower energies to the most probable states of the evolving system. In this study, we take a local entropy in the configuration space for the reinforcement and apply the algorithm to a number of easy and hard optimization problems. The reinforced algorithm performs better than the standard quantum annealing algorithm in the quantum search problem, where the optimal parameters behave very differently depending on the number of solutions. Moreover, the reinforcements can change the discontinuous phase transitions of the mean-field p-spin model ($p>2$) to a continuous transition. The algorithm's performance in the binary perceptron problem is also superior to that of the standard quantum annealing algorithm, which already works better than a classical simulated annealing algorithm.
\end{abstract}

\maketitle

\section{Introduction}\label{S0}
Finding the minimum-energy configurations of an energy function is essential for understanding the (near) optimal behavior of many physical, biological, and social systems \cite{QC-prep-2013,protein-scirep-2012,portfolio-qmachin-2019}. This is not, however, an easy task even for very restricted classes of systems like two-local energy functions, binary perceptrons, and two-player games \cite{osborne-rep-2012,baldassi-pnas-2007,daskalakis-siam-2009}. In fact, it seems very unlikely for a local algorithm (classical or quantum) to be able to get around the extensive energy barriers (or entropy barriers) of the solution space of large-scale frustrated systems \cite{bellitti-prr-2012}. One, of course expects to obtain better algorithms by exploiting some relevant global (or locally extended) information from the complex energy landscape of the problem. In a classical algorithm this information can be provided, for example, by the local entropy of solutions in the neighborhood of the system configuration \cite{baldassi-jstat-2016}. In a quantum annealing algorithm the information in the unitary evolution of the system can be used to suppress quantum transitions to the excited states in the annealing process \cite{B-jpa-2009,C-prl-2013,takahashi-pra-2017}.

In a quantum annealing (QA) algorithm, the system starts from the ground state of an easy initial Hamiltonian and undergoes an adiabatic evolution at zero temperature to reach the ground state of a hard final Hamiltonian \cite{Nishimori-pre-1998,Farhi-arxiv-2000,Das-rev-2008,Aharonov-sima-2008}. Importantly, there are different physical platforms which can be employed for practical implementation of quantum annealing processes \cite{qa-nature-2011,qa-nc-2013,QArydberg-natcom-2017,QAlattic-pra-2017,Lidar-revphys-2018,Lidar-nphys-2021}. The algorithm's performance is, however, limited by the nature of phase transitions that occur in the annealing process, especially close to the final Hamiltonian in a frustrated system \cite{AHJ-pnas-2010,qxor-prl-2010,qxor-pre-2011,qxor-pra-2012}. Nevertheless, there are problems where the quantum annealing algorithm is expected to work better than the classical annealing or Monte Carlo algorithms due to the structure of the solution space \cite{baldassi-pnas-2018,Hastings-arxiv-2020,Hatings-acm-2021}. This highlights the role of appropriate modifications in the energy landscape in enhancing the efficiency of an optimization algorithm.

The price that is paid for benefiting from global information in an algorithm is to work with nonlocal energy functions. Sometimes it is possible to map such a nonlocal function to a local one by introducing an additional set of auxiliary variables at the expense of increasing the computational costs of the original problem \cite{globalgame-2011,sign-prb-2012}. On the other hand, one may resort to effective theories, working with local approximations of the Hamiltonian \cite{localCA-pra-2014,localCA-pnas-2017}. The nature of energy terms that are added to the Hamiltonian is critical here for an efficient sampling of the optimal states. For instance, we know that non-stoquastic and even non-Hermitian Hamiltonians could enhance the efficiency of the standard quantum annealing algorithm, which usually works with a stoquastic and Hermitian Hamiltonian \cite{hormozi-prb-2017,susa-pra-2017,nesterov-pra-2012}. In addition, the way that such Hamiltonians are incorporated in an optimization algorithm plays an important role in its performance. This enters, for example, in the time dependence of the Hamiltonian parameters or the rate of changing the parameters, especially close to a phase transition \cite{ali-pra-2010,pspin-pra-2018,pspin-pra-2019,SG-njphys-2019,optimal-prl-2021,quench-prxq-2021}. Another strategy is to search for the optimal parameters, for example in a reinforcement learning algorithm, given the annealing schedule \cite{RQA-scirep-2020}.

\begin{figure}
\includegraphics[width=12cm]{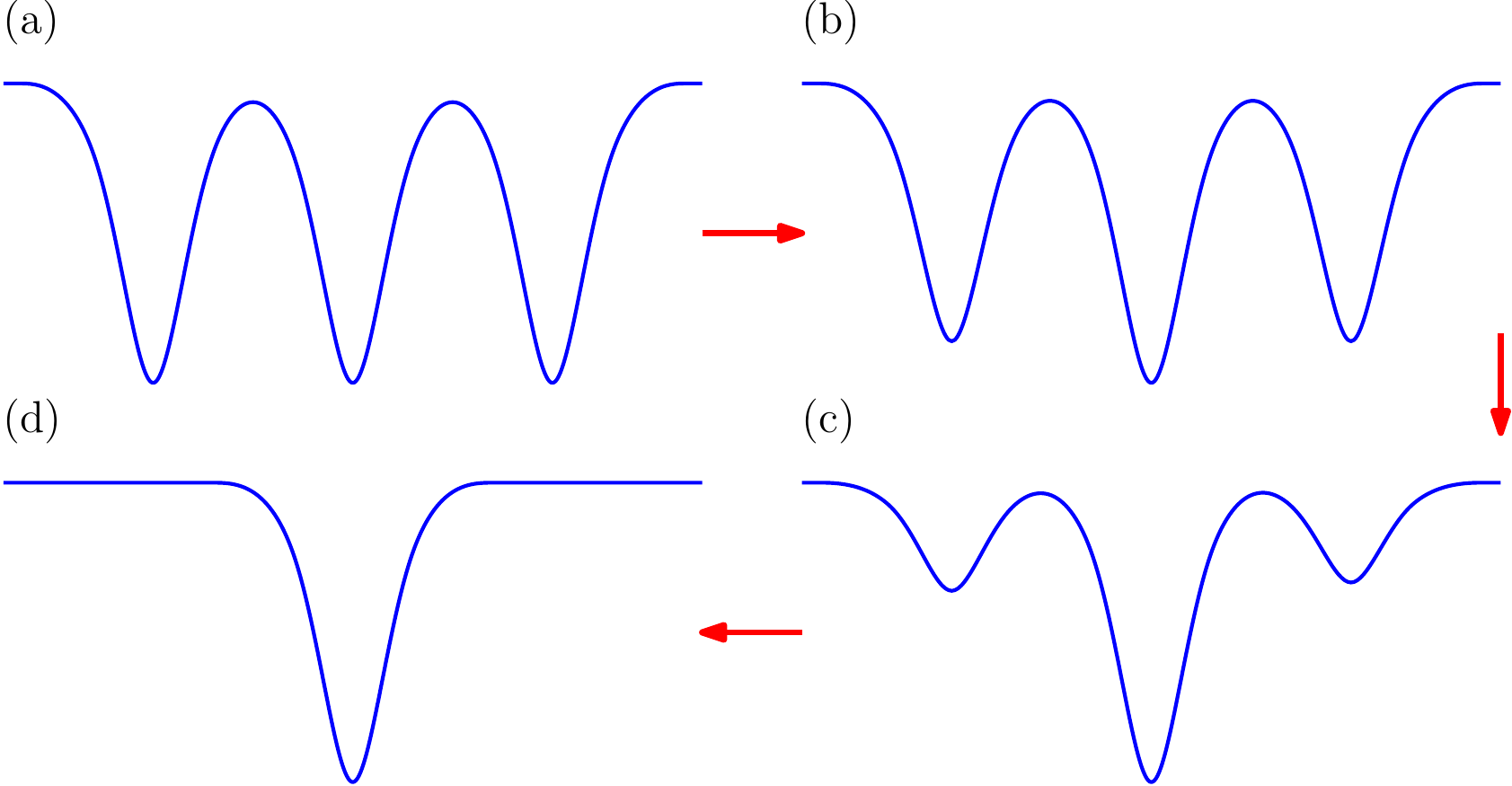} 
\caption{Changing the energy landscape by reinforcement. A small noise can break the problem degeneracy in case there are many solutions. Then reinforcement can be used to change slowly the energy landscape to have a nearly trivial one.}\label{f1}
\end{figure}

In this paper, we shall see how a simulated quantum annealing algorithm works in the presence of an entropic reinforcement. This study is an extension of the method introduced in \cite{QR-pra-2017,QR-pra-2018} which here is applied to different problems. The main idea is that reinforcement is able to increase the minimum energy gap in an annealing process and consequently to enhance the performance of a quantum annealing algorithm. The reason is that by reinforcement we are indeed changing the energy landscape (see Fig. \ref{f1}) to favor an optimal configuration \cite{SB-book-1998,BZ-prl-2006}. This is like increasing the energy gap between the optimal state and the other states of the system. We use the quantum state of the system to define a local entropy for each point in the configuration space which depends on the observation probabilities of the nearby configurations. In practice, this means that we need to estimate the quantum state during the annealing process by performing a continuous weak measurement of many copies of the system  \cite{lloyd-prl-1998,DK-pra-1999,lloyd-pra-2000,Qestimation-prl-2006,Qcontrol-book,MBQC-nature-2018}.

In the following, we report the results that are obtained by numerical simulations of the reinforced quantum annealing (rQA) for a number of prototypical optimization and search problems. We start with the study of the minimum energy gap in a random Ising model for a small number of spins. Then we take the quantum search problem and compare the results with the annealing version of the Grover algorithm. Next, we see how the reinforcement changes the nature of the phase transition in the mean-field p-spin model. Finally, we resort to a path-integral quantum Monte Carlo simulation to study the binary perceptron problem.

\section{Results}\label{S1}
Consider a classical spin system of $N$ binary variables $\boldsymbol\sigma=\{\sigma_i=\pm 1:i=1,\dots,N\}$, with energy function $E(\boldsymbol\sigma)$. This defines the problem Hamiltonian $\hat{H}_p=\sum_{\boldsymbol\sigma}E(\boldsymbol\sigma)|\boldsymbol\sigma\rangle\langle \boldsymbol\sigma|$ in the computational basis $|\boldsymbol\sigma\rangle$. The states $|\sigma_i\rangle$ are eigenstates of the $z$ component of the Pauli matrices $(\sigma_i^x,\sigma_i^y,\sigma_i^z)$. Hopping between the neighboring spin configurations is provided by $\hat{H}_0=-\Gamma\sum_i \sigma_{i}^x$. An interpolation between the two Hamiltonians is given by
\begin{align}
\hat{H}=(1-\tau)\hat{H}_0+\tau \hat{H}_p,
\end{align}
where $\tau \in [0,1]$. This can be considered the Hamiltonian of a quantum random walker in the energy landscape of the problem $E(\boldsymbol\sigma)$.

Reinforcement is to modify the Hamiltonian according to the quantum state of the system, for instance in proportion to $\ln|\psi_0(\boldsymbol\sigma)|^2$. In this way, we obtain the reinforced Hamiltonian \cite{QR-pra-2017},
\begin{align}\label{Hr0}
\hat{H}_r=(1-\tau)\hat{H}_0+\tau \hat{H}_p-r\sum_{\boldsymbol\sigma}\ln|\psi_0(\boldsymbol\sigma)|^2 |\boldsymbol\sigma\rangle\langle \boldsymbol\sigma|.
\end{align}
Here $\psi_0(\boldsymbol\sigma)$ is (an estimate of) the ground-state wave function and $r$ is the reinforcement parameter. Both the wave function $\psi_0$ and the parameter $r$ could in general depend on $\tau$. 
Note that for positive $r$ the more likely states in the wave function have lower energies. And the expectation value of the reinforcement term $-r\sum_{\boldsymbol\sigma}|\psi_0(\boldsymbol\sigma)|^2\ln|\psi_0(\boldsymbol\sigma)|^2$ is proportional to the entropy of the wave function in the given basis (not the von Neumann entropy). Thus, for very large and positive $r$ the reinforcement term favors states of zero entropy which are concentrated on a single spin configuration. On the other hand, for very negative $r$ the state of maximum entropy is favored which is the ground state of $\hat{H}_0$.

Note that the wave function can always be expanded in terms of many-body spin interactions $\psi_0(\boldsymbol\sigma)\propto \exp(\sum_iB_i\sigma_i/2+\sum_{(ij)}K_{ij}\sigma_i\sigma_j/2+\cdots)$. Therefore, the reinforcement term can be approximated by local Hamiltonians when higher-order correlations in the wave function are negligible \cite{QR-pra-2018}. In particular, if $\psi_0(\boldsymbol\sigma)$ is approximated by a product state, the reinforcement Hamiltonian is represented by one-spin interactions with local fields $B_i$ from the wave function. These fields then provide a bias towards the (instantaneous) ground state of the system that is very much like the bias fields that are introduced in Ref. \cite{QAbias-prl-2019}.    

Let us rewrite the reinforced Hamiltonian as
\begin{align}\label{Hr00}
\hat{H}_r=(1-\tau)\hat{H}_0+\tau \hat{H}_p+r\sum_{\boldsymbol\sigma}S(\boldsymbol\sigma)|\boldsymbol\sigma\rangle\langle \boldsymbol\sigma|,
\end{align}
with the reinforced entropy $S(\boldsymbol\sigma)= -\ln|\psi_0(\boldsymbol\sigma)|^2$.  
A generalization of the reinforcement entropy reads
\begin{align}\label{Sr0}
S_q(\boldsymbol\sigma:\lambda)=-\frac{1}{Q}\left(\left(\sum_{\boldsymbol\sigma'}e^{-\lambda D(\boldsymbol\sigma,\boldsymbol\sigma')}|\psi_0(\boldsymbol\sigma')|^2\right)^Q-1\right),
\end{align}
where $D(\boldsymbol\sigma,\boldsymbol\sigma')=\sum_i(\sigma_i-\sigma_i')^2$ is the Hamming distance of the two spin configurations. Note that for $\lambda\to \infty$ and $Q\to 0$ we recover the original entropy. Here the reinforcement term consists of a sum over all configurations that are close to the reference configuration $\boldsymbol\sigma$. The typical distance from the reference configuration is controlled by the Lagrange parameter $\lambda$. 
The parameter $\lambda$ determines the size of window that is used to compute the coarse grained probabilities in the entropy. Smaller values of $\lambda$ make the energy landscape a smoother function and help it to escape from the local minima. The optimal $\lambda$ depends on the typical size of the basins of attraction of the local minima, which is not \textit{a priori} known and should be tuned in each different problem. The parameter $Q$ determines the importance of the states of smaller probabilities compared to the most probable state(s). Note that positive and negative values of $Q$ both assign smaller energies to higher probabilities in proportion to $|\psi_0(\boldsymbol\sigma)|^{2|Q|}$ and $|\psi_0(\boldsymbol\sigma)|^{-2|Q|}$, respectively. Clearly, the reinforcement is more sensitive to changes in the probabilities for negative values of $Q$. The expectation values of the generalized entropy in state $|\psi_0\rangle$ are
\begin{align}
S_q(\lambda) &=\sum_{\boldsymbol\sigma}|\psi_0(\boldsymbol\sigma)|^2S_q(\boldsymbol\sigma:\lambda).
\end{align}

As an example, consider a mean-field approximation of the ground state, assuming a product state $\psi_0(\boldsymbol\sigma) \propto \exp(\sum_i B_i\sigma_i/2)$ with real parameters $B_i$. Then
\begin{align}
S_q(\boldsymbol\sigma:\lambda)=-\frac{1}{Q}\left(\prod_i\left(\frac{e^{B_i-\lambda(\sigma_i-1)^2}+e^{-B_i-\lambda(\sigma_i+1)^2}}{e^{B_i}+e^{-B_i}}\right)^Q-1\right).
\end{align}
And for the average entropy we get
\begin{align}
S_q(\lambda) &=-\frac{1}{Q}\left(\prod_i[\pi_q(m_i:\lambda)+\pi_q(-m_i:\lambda)]-1\right),
\end{align}
where
\begin{align}
\pi_q(m_i:\lambda) &=(\frac{1+m_i}{2})(\frac{1+m_i+e^{-4\lambda}(1-m_i)}{2})^Q.
\end{align}

Returning to Eq. \ref{Sr0}, let us rewrite the generalized entropy as 
\begin{align}
S_q(\boldsymbol\sigma:\lambda)=-\frac{1}{Q}\left( e^{Q \ln \langle e^{-\lambda D(\boldsymbol\sigma,\boldsymbol\sigma')}\rangle }-1\right).
\end{align}
Note that the expectations are taken with respect to the $\boldsymbol\sigma'$ variables, i.e., $\langle O\rangle=\sum_{\boldsymbol\sigma'}|\psi_0(\boldsymbol\sigma')|^2O(\boldsymbol\sigma')$.
Now, we may use the cumulant expansion to get
\begin{align}
\ln \langle e^{-\lambda D(\boldsymbol\sigma,\boldsymbol\sigma')}\rangle=\sum_{n=1}^{\infty}\frac{(-\lambda)^n}{n!}\langle D(\boldsymbol\sigma,\boldsymbol\sigma')^n\rangle_c,
\end{align}
where the connected expectation values are $\langle O^n\rangle_c=(\frac{\partial}{\partial\lambda})^n\ln \langle e^{-\lambda O}\rangle|_{\lambda=0}$. Thus
\begin{align}
S_q(\boldsymbol\sigma:\lambda)=-\frac{1}{Q}\left( e^{Q\sum_{n=1}^{\infty}\frac{(-\lambda)^n}{n!}\langle D(\boldsymbol\sigma,\boldsymbol\sigma')^n\rangle_c}-1\right).
\end{align}

In the following we set $Q=q/N$ to get extensive reinforcement entropy.  
In practice, one may consider only the first leading terms of the cumulant expansion. 
The limit $q\to 0$ then provides a local approximation of the entropy.
Finally, the reinforced Hamiltonian reads 
\begin{align}\label{Hr1}
\hat{H}_r=(1-\tau)\hat{H}_0+\tau \hat{H}_p+r\sum_{\boldsymbol\sigma}S_q(\boldsymbol\sigma:\lambda)|\boldsymbol\sigma\rangle\langle \boldsymbol\sigma|.
\end{align}
Note that the parameters $\tau(t), r(t), q(t)$, and $\lambda(t)$ could in general depend on real time $t$.

\begin{figure}
\includegraphics[width=16cm]{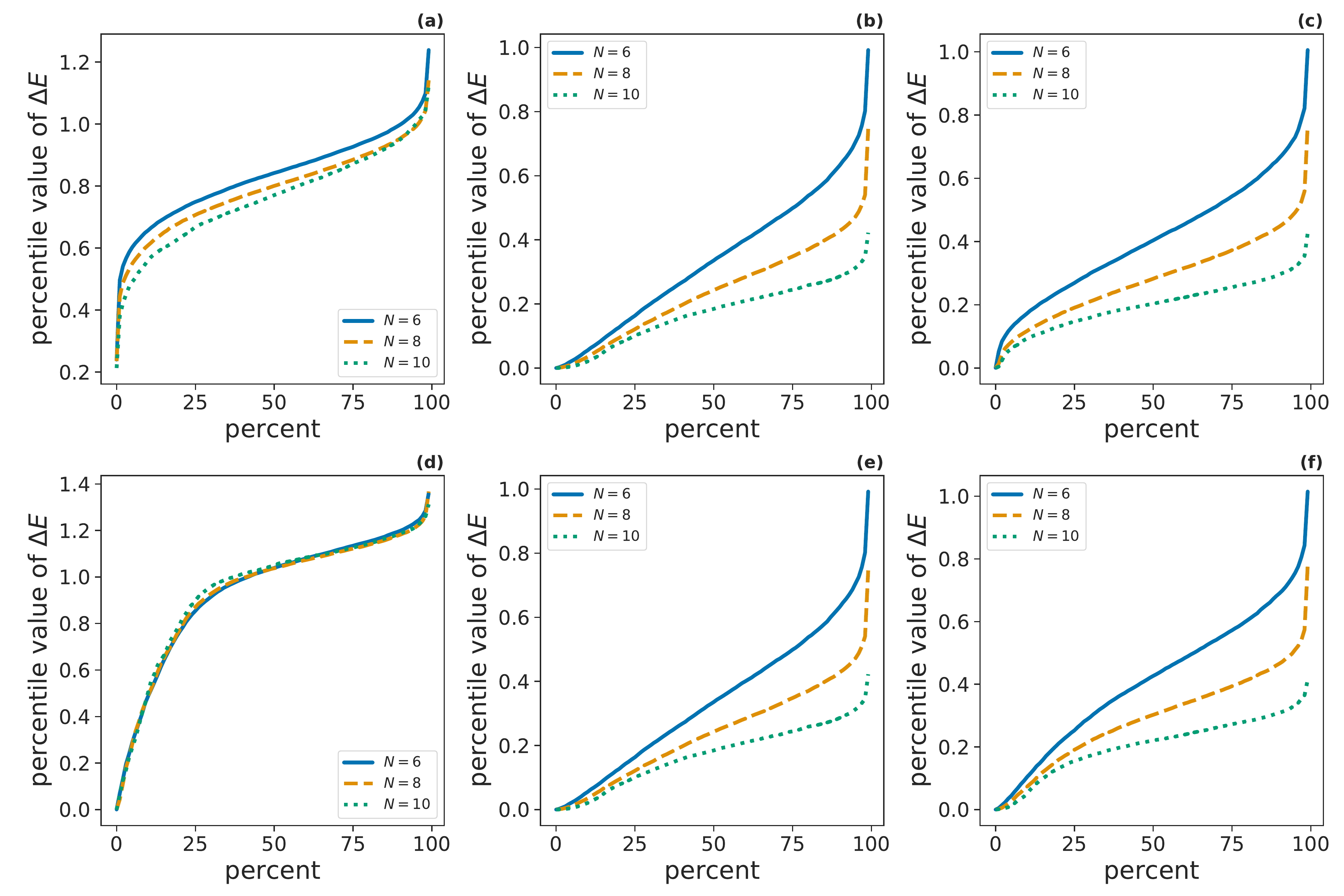} 
\caption{Statistics of the minimum gap $\Delta E$ of the random Ising model in the annealing process. The percentile value of $\Delta E$ is reported for two cases using the instantaneous ground state (top panels) or the Boltzmann weights (bottom) for reinforcement. The horizontal axis gives the percentage of problem instances that display a minimum energy gap smaller than $\Delta E$. The statistics is obtained for at least $2000$ independent and random realizations of the problem. The left, middle, and right panels show the cases of $q=-2$, no reinforcement, and $q=+2$, respectively. The results are obtained by following the exact ground state of the reinforced Hamiltonian for a small number of spins $N$. The annealing process starts with $r=\lambda=0$ at $\tau=0$ and the parameters increase linearly with the evolution time where $\delta \tau=0.001$, $\delta r=0.01$, and $\delta \lambda=0.01$. Here $\Gamma=1$ and we set $\beta=4$ in the Boltzmann weights. The numerical data are obtained with the linalg.eigh method of scientific python.}\label{f2}
\end{figure}

\subsection{Replacing the quantum expectations with thermal ones}\label{S11}
In the above formulation of reinforcement, we need to have an estimation of the expectation values of distances $D(\boldsymbol\sigma,\boldsymbol\sigma')^2$ with respect to the instantaneous ground state of the system. This, in principle, can be done by a weak measurement of many copies of the system in the annealing process \cite{Qestimation-prl-2006}. An alternative is to exploit the statistical information in the thermal Boltzmann weights of the system to avoid the quantum measurements. So let us replace the quantum expectations in the reinforcement entropy with thermal averages,
\begin{align}
S_q(\boldsymbol\sigma:\lambda)=-\frac{1}{Q}\left(\left(\sum_{\boldsymbol\sigma'} e^{-\lambda D(\boldsymbol\sigma',\boldsymbol\sigma)}|\phi(\boldsymbol\sigma')|^2\right)^Q-1\right),
\end{align}
where now
\begin{align}
|\phi(\boldsymbol\sigma')|^2=\frac{e^{-\beta E(\boldsymbol\sigma')}}{\sum_{\boldsymbol\sigma''}e^{-\beta E(\boldsymbol\sigma'')}}=e^{-\beta [E(\boldsymbol\sigma')-F(0)]}.
\end{align}
Here $F(0)$ is the free energy of the thermal system at inverse temperature $\beta$.

Figure \ref{f2} shows the statistics of the minimum gap for a fully connected random Ising model in the annealing process. The energy function for this system is
\begin{align}
E(\boldsymbol\sigma)=-\sum_{i<j}\frac{J_{ij}}{\sqrt{N}}\sigma_i\sigma_j-\sum_{i}h_{i}\sigma_i,
\end{align}
where the fields $h_i$ and couplings $J_{ij}$ are random variables drawn from a normal distribution of mean $0$ and variance $1$ \cite{SM}. 
The figure compares two cases of using the instantaneous ground state or the Boltzmann weights for reinforcement. The results are obtained by finding the exact ground state of the reinforced Hamiltonian for a small number of spins $N=6,8,10$. The figure shows the percentage or fraction of problem instances that display a minimum energy gap smaller than $\Delta E$. We see that the minimum gap is significantly larger when we have reinforcement with a negative $q$ compared to cases of no reinforcement or reinforcement with a positive $q$. Moreover, the scaling of the minimum gap with the size of the system is better when we use the Boltzmann weights for reinforcement. In practice, however, we do not have access to the exact free energy of large systems. Thus, one has to resort to an approximate free energy, e.g., estimated by the Bethe approximation (see Appendix \ref{app1}). In the following, we investigate the effect of reinforcement in other problems of different characters and sizes. In all the studied problems we focus mainly on the results which are obtained by using the quantum expectations. Examples of using the thermal expectations are to show that the method also works in this case to get around the practical difficulties of continuous quantum measurements.

\subsection{The adiabatic quantum search problem}\label{S12}
Consider a search space of size $2^N$ partitioned into the subspaces of solutions ($|g\rangle$) and excited states ($|e\rangle$). An arbitrary state of the system is represented by 
\begin{align}
|\psi(t)\rangle=\alpha(t)|g\rangle+\beta(t)|e\rangle.
\end{align}   
The initial state is the ground state of $H_0=\mathsf{1}-|\psi(0)\rangle\langle \psi(0)|$ and the solutions are the ground states of $H_p=\mathsf{1}-|g\rangle\langle g|$ with zero energies \cite{fdsearch-prl-2005,fdsearch-pra-2017,search-pra-2020}. Here the reinforced Hamiltonian reads
\begin{align}
\hat{H}_r=(1-\tau)\hat{H}_0+\tau \hat{H}_p+rS_q(|\alpha(t)|^2)|g\rangle\langle g|+rS_q(|\beta(t)|^2)|e\rangle\langle e|,
\end{align}
where $S_q(x)=-(x^q-1)/q$. 

The Schrodinger equation gives the time evolution of the state (for $\hbar=1$),
\begin{align}
\hat{i}\frac{d\alpha(t)}{dt}=(1-\tau) [1-P(0)] \alpha(t)-(1-\tau) \sqrt{P(0)[1-P(0)]}\beta(t)+r\alpha(t)S_q(|\alpha(t)|^2),\\
\hat{i}\frac{d\beta(t)}{dt}=[\tau+(1-\tau)P(0)]\beta(t)-(1-\tau) \sqrt{P(0)[1-P(0)]}\alpha(t)+r\beta(t)S_q(|\beta(t)|^2),
\end{align}   
where $P(t)=|\langle g|\psi(t) \rangle|^2$ is the success probability in a measurement.

\begin{figure}
\includegraphics[width=16cm]{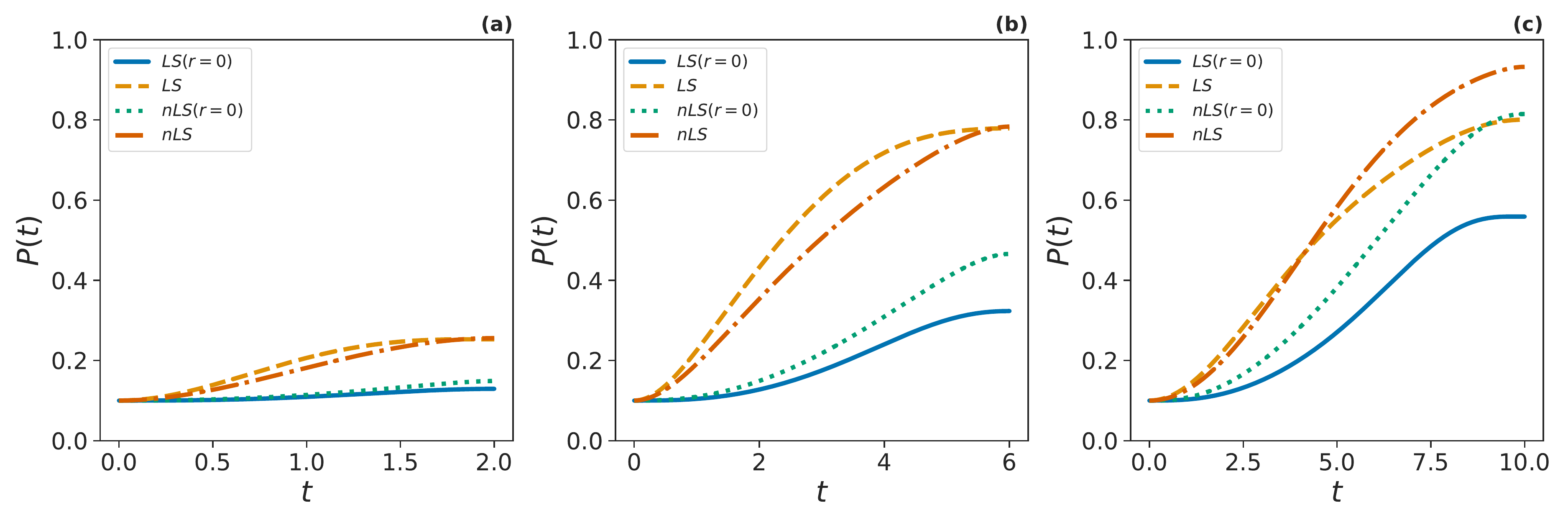} 
\caption{Time evolution of the success probability in the reinforced quantum search problem. The results for the linear schedule (LS) and nonlinear schedule (nLS) are given for two cases of with and without reinforcement. The parameters in the reinforced schedules are: (a) for $T=2: LS(q=0.01,r=-1), nLS(q=0.01,r=-0.75)$, (b) for $T=6: LS(q=0.01,r=-0.85), nLS(q=0.01,r=-0.65)$, and (c) for $T=10: LS(q=1.5,r=-0.75), nLS(q=1.5,r=-0.5)$. The numerical data are obtained with the integrate.odeint method of scientific python.}\label{f3}
\end{figure}

\begin{figure}
\includegraphics[width=16cm]{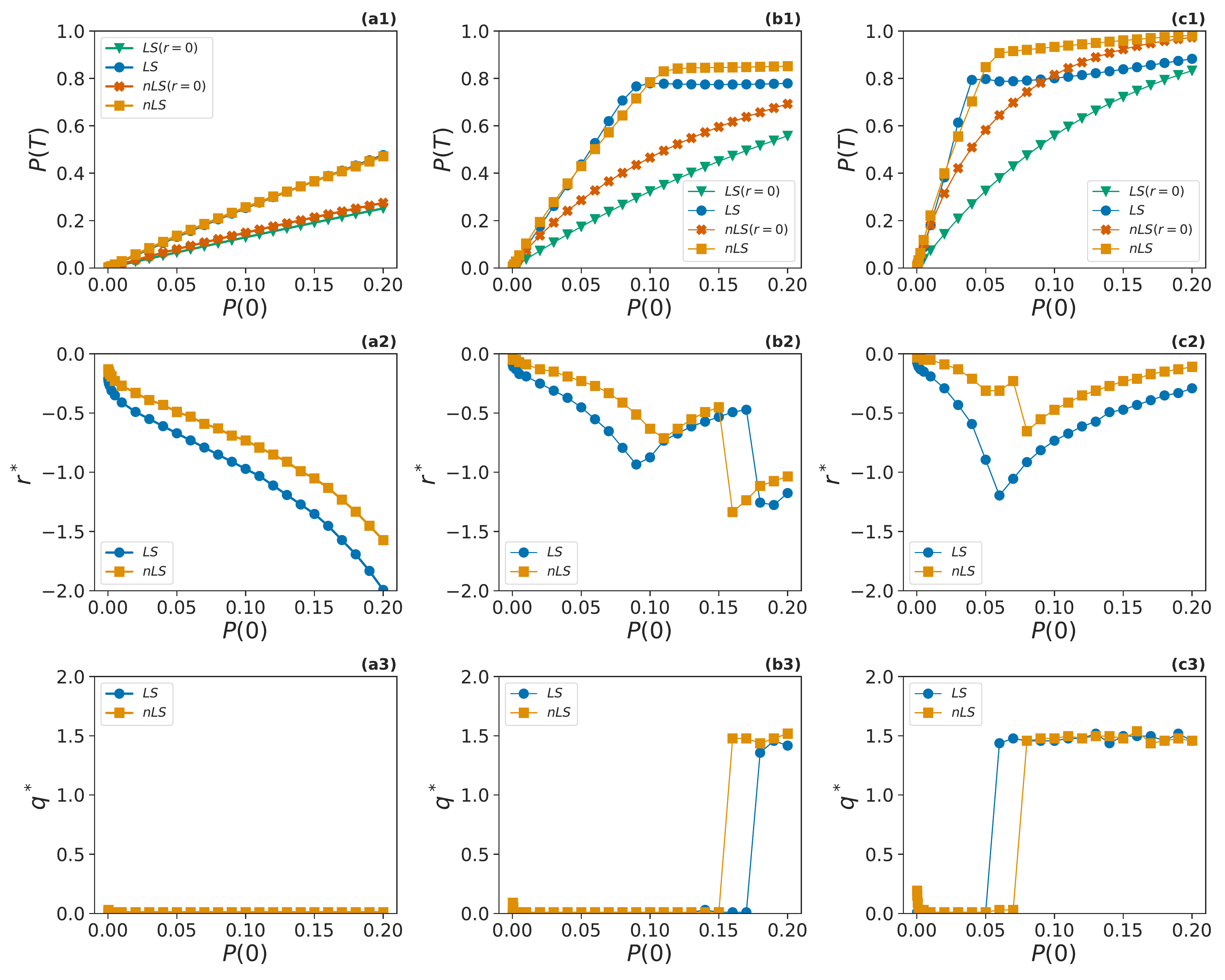} 
\caption{The success probability and optimal parameters in the reinforced quantum search problem. The results for the linear schedule (LS) and nonlinear schedule (nLS) are given for cases with and without reinforcement. The success probability $P(T)$ at the end of the process and the optimal parameters $(r^*, q^*)$ are plotted vs $P(0)$ for evolution times $T=2$ (panels (a1),(a2),(a3)), $T=5$ (panels (b1),(b2),(b3)), and $T=10$ (panels (c1),(c2),(c3)). The parameters in the reinforced schedules are restricted to $r\in (-2,0)$ and $q \in (0,2)$. The numerical data are obtained with the integrate.odeint method of scientific python.}\label{f4}
\end{figure}

Figure \ref{f3} shows how the success probability changes with time for a specific $P(0)=0.1$. We are using different annealing schedules \cite{fdsearch-pra-2017}: the linear schedule with $\tau(t)=t/T$ and the nonlinear schedule, where
\begin{align}
\tau(t) &=\frac{1}{2}\left(1-\sqrt{\frac{P(0)}{1-P(0)}}\tan\left((1-2\frac{t}{T})\phi\right)\right),\\
\phi &=\arctan\left(\sqrt{\frac{1-P(0)}{P(0)}}\right).
\end{align}   
The latter is guaranteed to display the Grover scaling of the evolution time with the size of the solution space $T \propto 1/\sqrt{P(0)}$ \cite{fdsearch-pra-2017}. The behaviors are compared for the cases with and without reinforcement.
For the sake of simplicity, we assume that the reinforcement parameter is fixed to a negative value to favor the solution space. We see in the figure that adding reinforcement can significantly improve the success probability for both the schedules.

In Fig. \ref{f4} we report the success probability at the end of the annealing process along with the optimal parameters $r^*$ and $q^*$, which maximize this probability. More precisely, we restricted the parameters to $r\in (-2,0)$ and $q \in (0,2)$. Interestingly, the reinforced Hamiltonian displays two distinct phases of easy and hard regimes depending on $P(0)$ and the evolution time $T$. The two phases are easier to distinguish by looking at the behavior of $q^*$, which changes abruptly from $q^*\simeq 1.5$ in the easy phase [$P(0)>P_c(T)$] to $q^*\simeq 0$ in the hard phase [$P(0)<P_c(T)$]. From the Grover scaling we expect to have $P_c(T)\propto 1/T^2$. In our numerical experiments, we could actually observe a greater success probability in the hard phase when the parameter $q$ takes very negative values with very small, but negative, reinforcements. 
In the following sections, we shall see other examples that support this observation.

\begin{figure}
\includegraphics[width=16cm]{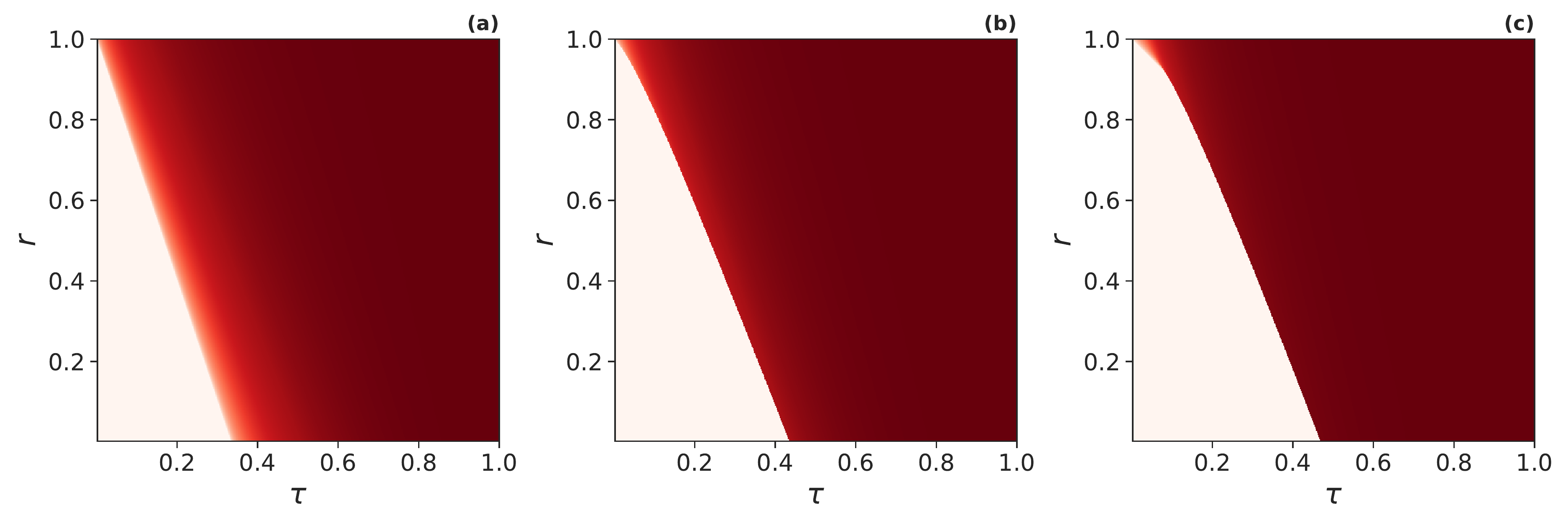} 
\caption{The ground-state magnetization of the mean-field p-spin model (a) $p=2$, (b) $p=3$, and (c) $p=5$. The magnetization in the thermodynamic limit is given in terms of the reinforcement parameter $r$ and the evolution time $\tau$ of the annealing process. The results are obtained by minimizing the expectation value of the reinforced Hamiltonian for a product state when $Q=0$.}\label{f5}
\end{figure}

\begin{figure}
\includegraphics[width=16cm]{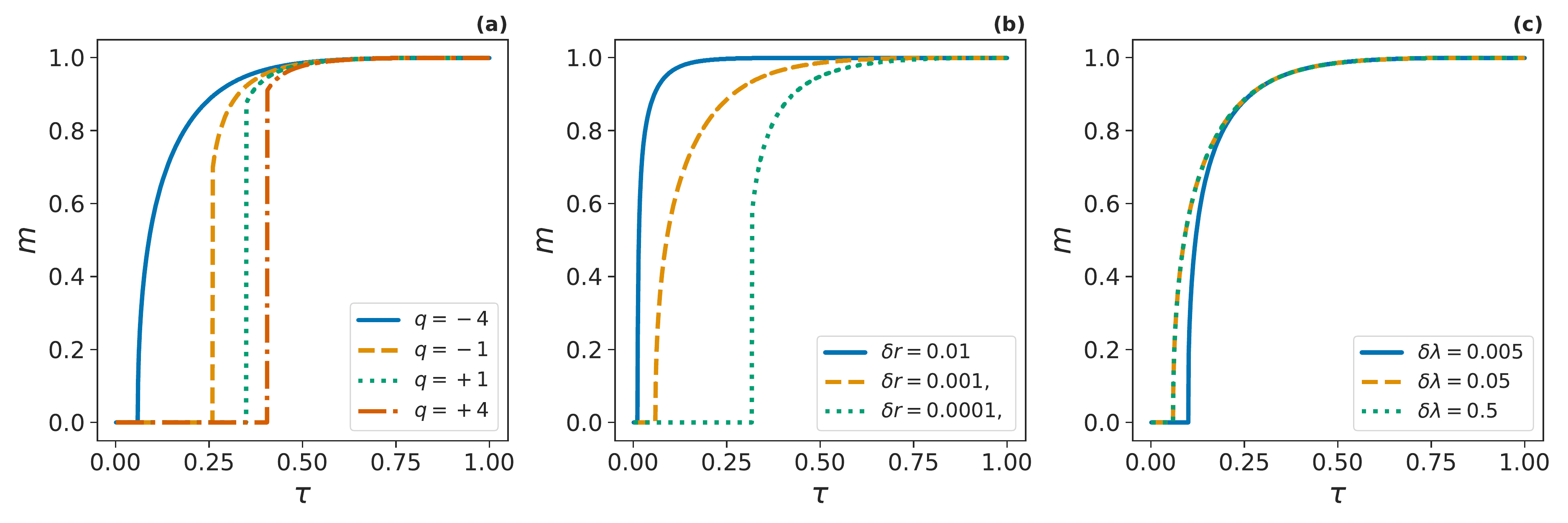} 
\caption{Quantum annealing of the mean-field p-spin model for $p=3$. The ground state magnetization $m$ in the thermodynamic limit is plotted vs the evolution time $\tau$. The annealing process starts with $r=\lambda=0$ at $\tau=0$ and the parameters increase linearly with the evolution time: $r(\tau+\delta\tau)=r(\tau)+\delta r$, and $\lambda(\tau+\delta\tau)=\lambda(\tau)+\delta \lambda$. The panels compare different cases of the parameter values: (a) $\delta \tau=0.001,\delta r=0.001, \delta \lambda=0.05$, (b) $q=-4, \delta \tau=0.001,\delta \lambda=0.05$, and (c) $q=-4, \delta \tau=0.001,\delta r=0.001$. The results are obtained by minimizing the expectation value of the reinforced Hamiltonian for a product state.}\label{f6}
\end{figure}

\subsection{The mean-field p-spin model}\label{S13}
As another example we consider the mean-field p-spin model with the following energy function
\begin{align}
E(\boldsymbol\sigma)=-N(\frac{1}{N}\sum_i \sigma_i)^p.
\end{align}
This problem has trivial ground states but the standard quantum annealing is not efficient for $p\ge 3$ due to a first-order transition with an exponentially small energy gap in the annealing process \cite{JKK-epl-2010,BS-jstat-2012,FDV-pra-2011,SN-jphis-2012}. Here the annealing Hamiltonian is $\hat{H}=\tau \hat{H}_p+(1-\tau)\hat{H}_0$ with $\hat{H}_p=\sum_{\boldsymbol\sigma}E(\boldsymbol\sigma)|\boldsymbol\sigma\rangle\langle \boldsymbol\sigma|$ and $\hat{H}_0=-\sum_i\sigma_i^x$, where we set $\Gamma=1$. This Hamiltonian is symmetric under permutation of the spins and by the quantum de Finetti theorem \cite{BH-acm-2013,cirac-pra-2013} the ground state is well approximated by a product state for $N\to \infty$, 
\begin{align}
\psi_0(\boldsymbol\sigma)= \prod_i \left(\frac{e^{B\sigma_{i}/2}}{\sqrt{2\cosh B}}\right).
\end{align}
Thus the expectation value of the Hamiltonian is given by
\begin{align}
\frac{\langle \psi_0|\hat{H}|\psi_0\rangle}{N}=-\tau m^p-(1-\tau)\sqrt{1-m^2},
\end{align}
with $m=\tanh(B)$. 

Like before the reinforced Hamiltonian reads
\begin{align}\label{Hr2}
\hat{H}_r=\hat{H}+r\sum_{\boldsymbol\sigma}S_q(\boldsymbol\sigma:\lambda)|\boldsymbol\sigma\rangle\langle \boldsymbol\sigma|.
\end{align}
For $Q=0$, we get,
\begin{align}
\frac{\langle \psi_0|\hat{H}_r|\psi_0\rangle}{N}=[-\tau m^p+rs_0(m)]-(1-\tau)\sqrt{1-m^2},
\end{align}
where $s_0(m)=-\sum_{\boldsymbol\sigma}|\psi_0(\boldsymbol\sigma)|^2\ln|\psi_0(\boldsymbol\sigma)|^2$, that is
\begin{align}
s_0(m)=-\frac{1+m}{2}\ln(\frac{1+m}{2})-\frac{1-m}{2}\ln(\frac{1-m}{2}).
\end{align}
Figure \ref{f5} displays the ground-state magnetization as a function of the reinforcement parameter $r$ and $\tau \in [0,1]$.
We see how the reinforcement favors the minimum-energy configuration of the original problem because for larger $r$ the solution appears for smaller values of $\tau$.

For an arbitrary $Q$, the expectation value of the reinforced Hamiltonian is given by
\begin{align}
\frac{\langle \psi_0|\hat{H}_r|\psi_0\rangle}{N}=[-\tau m^p+r s_q(m:\lambda)]-(1-\tau) \sqrt{1-m^2},
\end{align}
where
\begin{align}
s_q(m:\lambda)=-\frac{1}{NQ}\left\{[\pi_q(m:\lambda)+\pi_q(-m:\lambda)]^N-1\right\}.
\end{align}
In the thermodynamic limit with $Q=q/N$, we get 
\begin{align}
s_q(m:\lambda)=-\frac{1}{q}\left(e^{-qs_0(m:\lambda)}-1\right),
\end{align}
where now
\begin{align}
s_0(m:\lambda)=-\frac{1+m}{2}\ln\frac{(1+m)+e^{-4\lambda}(1-m)}{2}-\frac{1-m}{2}\ln\frac{(1-m)+e^{-4\lambda}(1+m)}{2}.
\end{align}

\begin{figure}[H]
\center
\includegraphics[width=12cm]{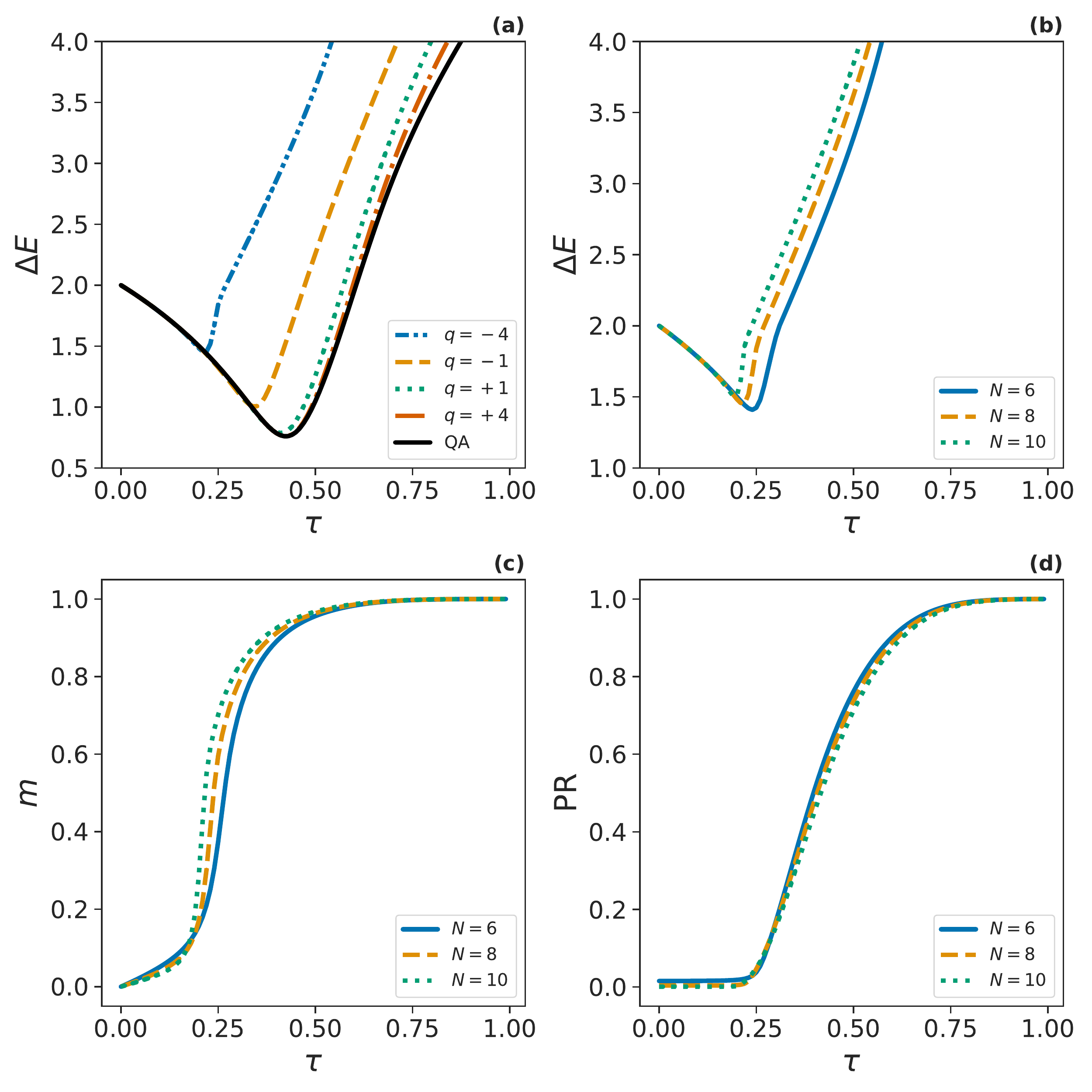} 
\caption{Quantum annealing of the mean-field p-spin model using the instantaneous ground state for reinforcement. The energy gap $\Delta E$ (panel (a) for $N=8$, panel (b) for $q=-4$), magnetization $m$ (panel (c) for $q=-4$), and participation ratio $PR=\sum_{\boldsymbol\sigma}|\psi_0(\boldsymbol\sigma)|^4$ (panel (d) for $q=-4$) are plotted for $p=3$ and finite sizes $N$. The results are obtained by following the exact ground state of the reinforced Hamiltonian for a small number of spins $N$. The annealing process starts with $r=\lambda=0$ at $\tau=0$ and the parameters increase linearly with the evolution time where $\delta \tau=0.001$, $\delta r=0.01$, and $\delta \lambda=0.01$. The numerical data are obtained with the linalg.eigh method of scientific python.}\label{f7}
\end{figure}

\begin{figure}[H]
\center
\includegraphics[width=12cm]{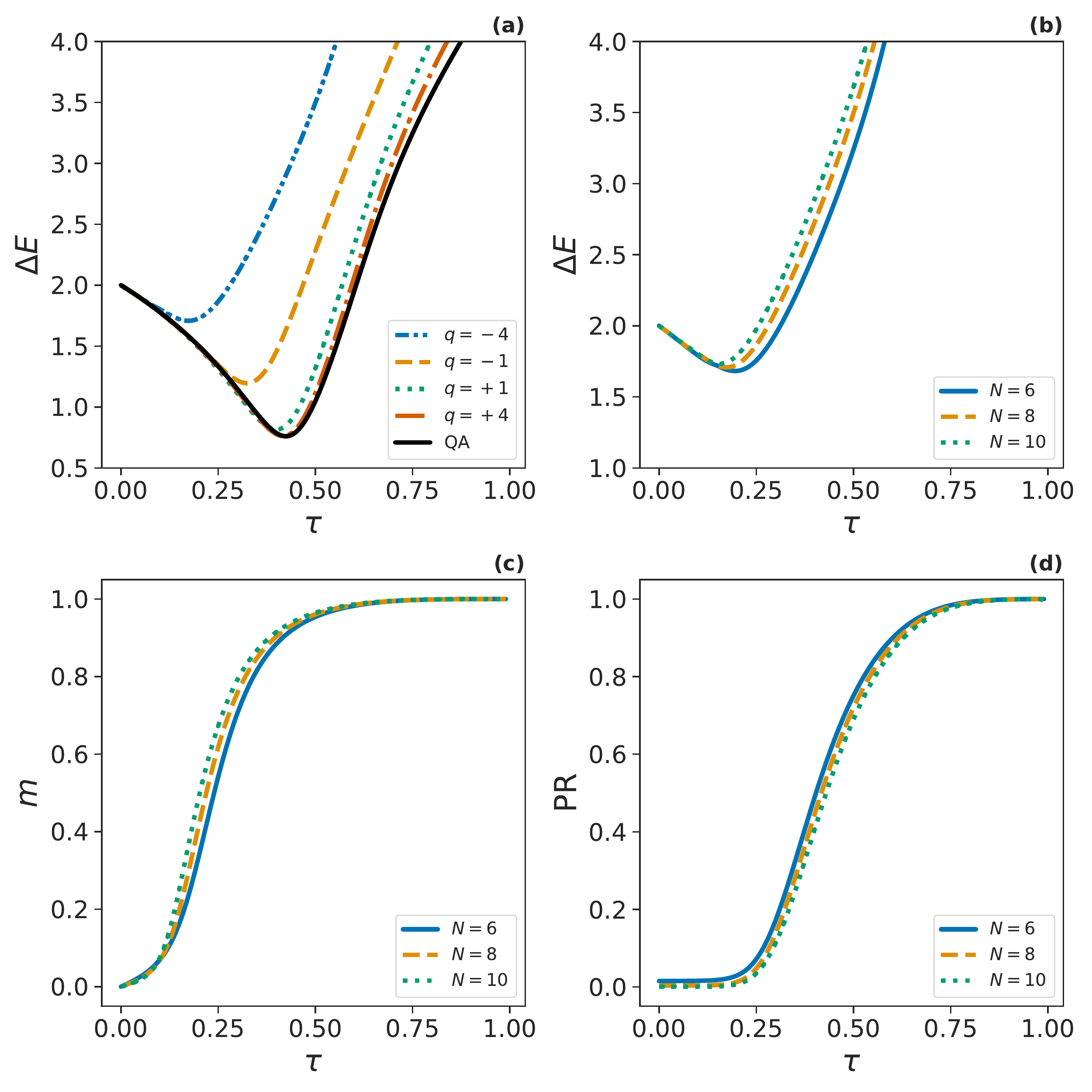} 
\caption{Quantum annealing of the mean-field p-spin model using the Boltzmann weights for reinforcement. The energy gap $\Delta E$ (panel (a) for $N=8$, panel (b) for $q=-4$), magnetization $m$ (panel (c) for $q=-4$), and participation ratio $PR=\sum_{\boldsymbol\sigma}|\psi_0(\boldsymbol\sigma)|^4$ (panel (d) for $q=-4$) are plotted for $p=3$. The results are obtained by following the exact ground state of the reinforced Hamiltonian for a small number of spins $N$. The annealing process starts with $r=\lambda=0$ at $\tau=0$ and the parameters increase linearly with the evolution time where $\delta \tau=0.001$, $\delta r=0.01$, and $\delta \lambda=0.01$. Here we set $\beta=4$ in the Boltzmann weights. The numerical data are obtained with the linalg.eigh method of scientific python.}\label{f8}
\end{figure}

In Fig. \ref{f6}, we observe the magnetization $m$ of the instantaneous ground state in the thermodynamic limit.
The results are obtained by minimizing the expectation value of the reinforced Hamiltonian with respect to a product state for $p=3$. As the figure shows, the nature of transition from the paramagnetic to ferromagnetic phase is different for positive and negative $q$. In the latter case we have a continuous phase transition which signals a change in the scaling of the minimum energy gap with the size of the system. Figures \ref{f7} and \ref{f8} show the behavior of the energy gap when the instantaneous ground state and the Boltzmann weight are used for reinforcement. The results are obtained by exact numerical simulation of the annealing process for small problem sizes. Again, we see that negative values of $q$ are more effective in increasing the minimum energy gap of this system.

\begin{figure}
\includegraphics[width=14cm]{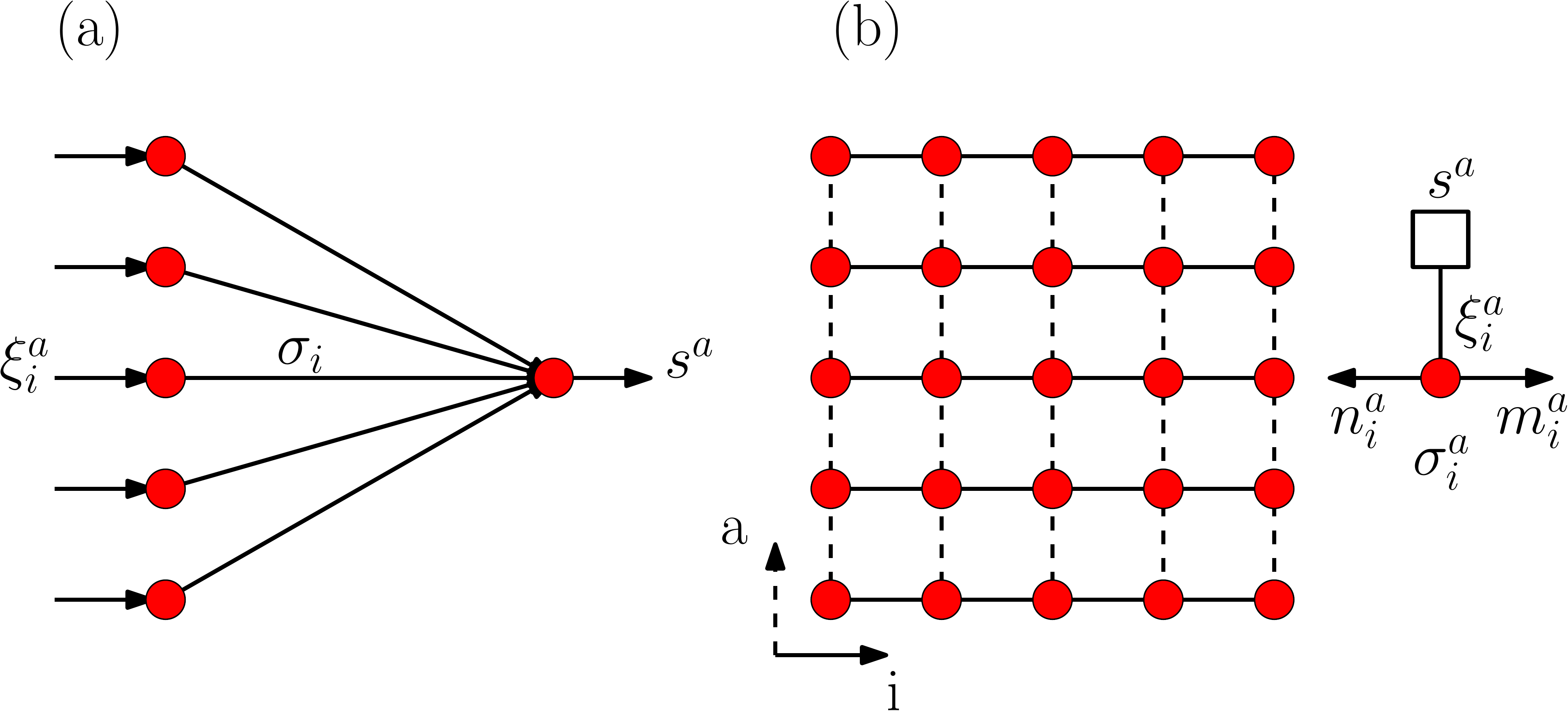} 
\caption{(a) The simple perceptron with one layer of input neurons $\xi_i$ and a single output neuron $s$. (b) A local representation of the problem for patterns $a=1,\dots,M$ along with the auxiliary variables $m_i^a$ and $n_i^a$.}\label{f9}
\end{figure}

\subsection{The perceptron problem: Quantum Monte Carlo simulations}\label{S14}
In this section, we consider the binary perceptron problem to check the algorithm's performance in a computationally  challenging problem \cite{R-report-1957,KM-jphys-1989,baldassi-pnas-2018,aubin-jphys-2019,Baldassi-arxiv-2021,cui-ML-2021}.
In a simple binary perceptron we have $N$ artificial neurons of states $\boldsymbol\xi=\{\xi_i=\pm 1:i=1\dots,N\}$ connected to a single output neuron of state $s=\mathrm{sgn}(\sum_i \sigma_i \xi_i)$ with binary weights $\boldsymbol\sigma=\{\sigma_i=\pm 1,i=1,\dots,N\}$. In a supervised learning of the perceptron we are to find the optimal weights $\boldsymbol\sigma^*$ to store a set of $M$ input-output patterns $\{\boldsymbol\xi^a,s^a: a=1,\dots, M\}$ such that $s^a=\mathrm{sgn}(\sum_i \sigma_i^* \xi_i^a)$. In the following, we consider $M$ random and independent patterns with equal probabilities for the $\pm 1$ values of $\xi_i^a$ and $s^a$. It is known that in this case one can store up to $M_c \simeq 0.83 N$ random patterns with an exponentially large number of isolated solutions $\boldsymbol\sigma^*$ in the space of weights \cite{KM-jphys-1989,aubin-jphys-2019,cui-ML-2021}. Because of the complex energy landscape of the problem, it is difficult for a classical simulated annealing algorithm to find a solution for large $N$ as the number of patterns approaches $M_c$. However, there are rare regions of close solutions in the configuration space which can be found by a standard quantum annealing algorithm \cite{baldassi-pnas-2018,Baldassi-arxiv-2021}. Therefore, in the following we compare the standard quantum annealing with a reinforced quantum walk where 
\begin{align}\label{Hr3}
\hat{H}_r=\hat{H}_0+\hat{H}_p+r(t)\sum_{\boldsymbol\sigma}S_q(\boldsymbol\sigma:\lambda)|\boldsymbol\sigma\rangle\langle \boldsymbol\sigma|,
\end{align}
that is, only the reinforcement parameter increases with time. Here $\hat{H}_0=-\Gamma \sum_i \sigma_i^x$ and $\hat{H}_p=\sum_{\boldsymbol\sigma}E(\boldsymbol\sigma)|\boldsymbol\sigma\rangle\langle \boldsymbol\sigma|$.     
The energy function in this problem just counts the number of unsatisfied patterns,
\begin{align}
E(\boldsymbol\sigma)=\sum_{a=1}^M [1-s^a\mathrm{sgn}(\sum_i\sigma_i\xi_i^a)].
\end{align}
Note that this is not a local energy because of the sign function. A local version of the above energy is given in Appendix \ref{app2} by introducing a set of auxiliary variables, see Fig. \ref{f9}.

To study the algorithm's performance for large problem sizes, in this section we resort to an approximation of the annealing process. The idea is to simulate the process by following the small changes in the Hamiltonian of the quantum system assuming that it is always in thermal equilibrium at a sufficiently small temperature. The partition function of such a quantum system can be mapped by the Suzuki-Trotter transformation to the partition function of a classical system. Then the standard Monte Carlo algorithm is used to obtain the average values of quantities like the energy or magnetization of the quantum system \cite{tosatti-prb-2002,pathMC-prb-2008}. 

More precisely, the partition function of the quantum problem with the reinforced Hamiltonian $\hat{H}_r$ at inverse-temperature $\beta$ is mapped to a classical problem of $P$ interacting replicas of the original system, 
\begin{align}
Z=\mathrm{Tr}e^{-\beta \hat{H}_r}=\lim_{P\to \infty}\sum_{\{\boldsymbol\sigma_{1},\dots,\boldsymbol\sigma_{P}\}}e^{-\beta\mathcal{E}(\vec{\boldsymbol\sigma})}.
\end{align}
The energy function of the replicated system $\vec{\boldsymbol\sigma}=\{\boldsymbol\sigma_1,\cdots,\boldsymbol\sigma_P\}$ reads,
\begin{align}
\beta\mathcal{E}(\vec{\boldsymbol\sigma})= \frac{\beta}{P}\sum_{\alpha=1}^P E_r(\boldsymbol\sigma_{\alpha})-\sum_{\alpha}\sum_{i}J_{\alpha}\sigma_{\alpha,i}\sigma_{\alpha+1,i}.
\end{align}
Here $2J_{\alpha}=\ln(\coth(\frac{\beta \Gamma}{P}))$ and $\boldsymbol\sigma_{P+1}=\boldsymbol\sigma_1$. The reinforced energy function is
\begin{align}
E_r(\boldsymbol\sigma_{\alpha})=E(\boldsymbol\sigma_{\alpha})+rS_q(\boldsymbol\sigma_{\alpha}:\lambda).
\end{align}

\begin{figure}
\includegraphics[width=14cm]{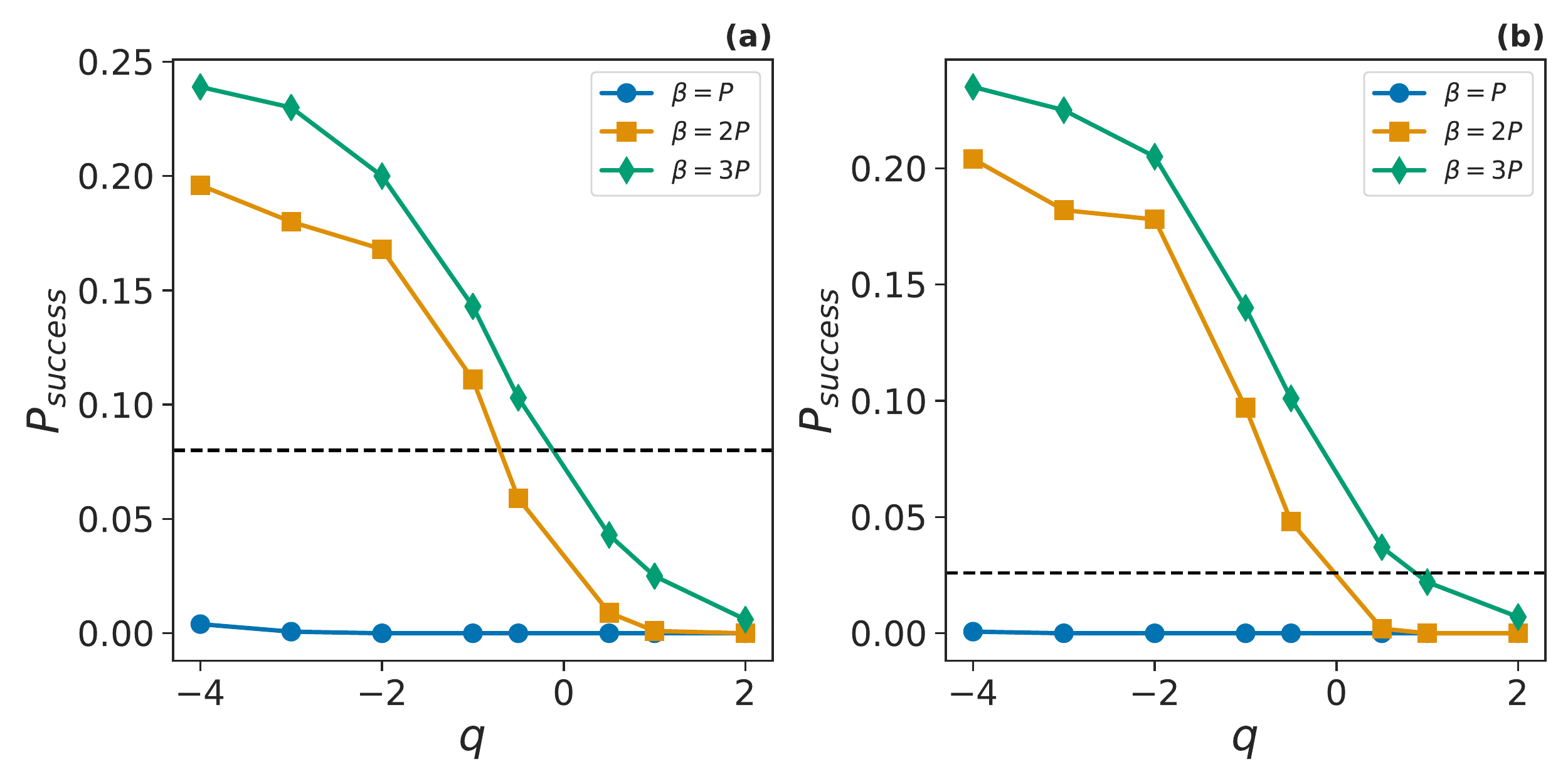} 
\caption{Success probability of the reinforced quantum annealing algorithm in the binary perceptron problem. The annealing process is simulated by the quantum Monte Carlo algorithm for (a) $\Gamma=1$ and (b) $\Gamma=2$. Each data point gives the fraction of successful runs $P_{success}$ in $10^3$ random instances of the problem with $N=200$ and $M=0.8 N$. The horizontal lines show the values we obtain by the standard quantum annealing algorithm. The parameters here are the number of replicas in the QMC algorithm $P=100$, the number of time steps $T=100$, the inverse temperature $\beta$, the equilibration time $\Delta t=10$, the rate of increasing the reinforcement parameter $\delta r=0.005$, and parameters of the generalized entropy $(q,\lambda)$. The value of $\lambda=0.5$ is fixed.}\label{f10}
\end{figure}

Here we consider the mean-field approximation of the generalized entropy with the scaling $Q=q/N$, 
\begin{align}
S_q(\boldsymbol\sigma:\lambda)=-\frac{N}{q}\left(\exp\left(\frac{q}{N}\sum_i\ln\left(\frac{e^{B_i-\lambda(\sigma_i-1)^2}+e^{-B_i-\lambda(\sigma_i+1)^2}}{e^{B_i}+e^{-B_i}}\right)\right)-1\right).
\end{align}
$B_i=\ln((1+m_i)/(1-m_i))$ are related to the local averages $m_i=\sum_{\alpha} \sigma_{\alpha,i}/P$. 
The annealing process starts at time step $t=0$ with random initial configurations for all replicas. Then for $t=1,\dots,T$ we do $\Delta t$ Monte Carlo sweeps to update the $\boldsymbol\sigma_{\alpha}$ configurations. The total number of spin flips is given by  $T\Delta t PN$.
At any time step $t$ we update the $B_i$ parameters and compute the minimum energy among the replicas. If the minimum energy is zero we have a successful process. 

\begin{figure}
\includegraphics[width=16cm]{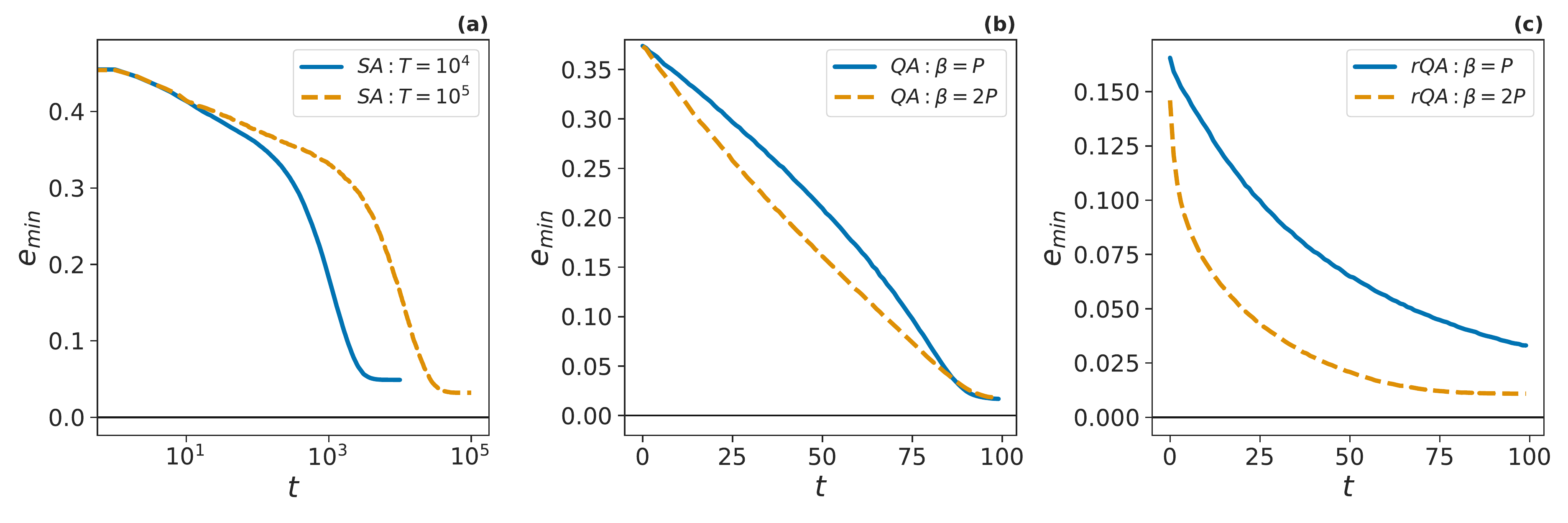} 
\caption{Time evolution of the average minimum energy in the QMC simulation of the classical and quantum annealing algorithms. We report the minimum-energy density obtained with (a) classical simulated annealing (SA), (b) standard quantum annealing (QA), and (c) reinforced quantum annealing (rQA) algorithms.  The data are the result of averaging over $10^3$ instances of the perceptron problem with $N=200$ and $M=0.8 N$. The other parameters in the SA algorithm are $\beta_0=0.01, \Delta t=10$ and $\delta\beta=10^{-3},10^{-4}$ for $T=10^{4},10^{5}$, respectively. In the QA algorithms $T=100$, $\Delta t=10$, $P=100$, and $\Gamma=1$. In the rQA algorithm $q=-4$, $\lambda=0.5$, and $\delta r=0.005$.}\label{f11}
\end{figure}

\begin{figure}
\includegraphics[width=12cm]{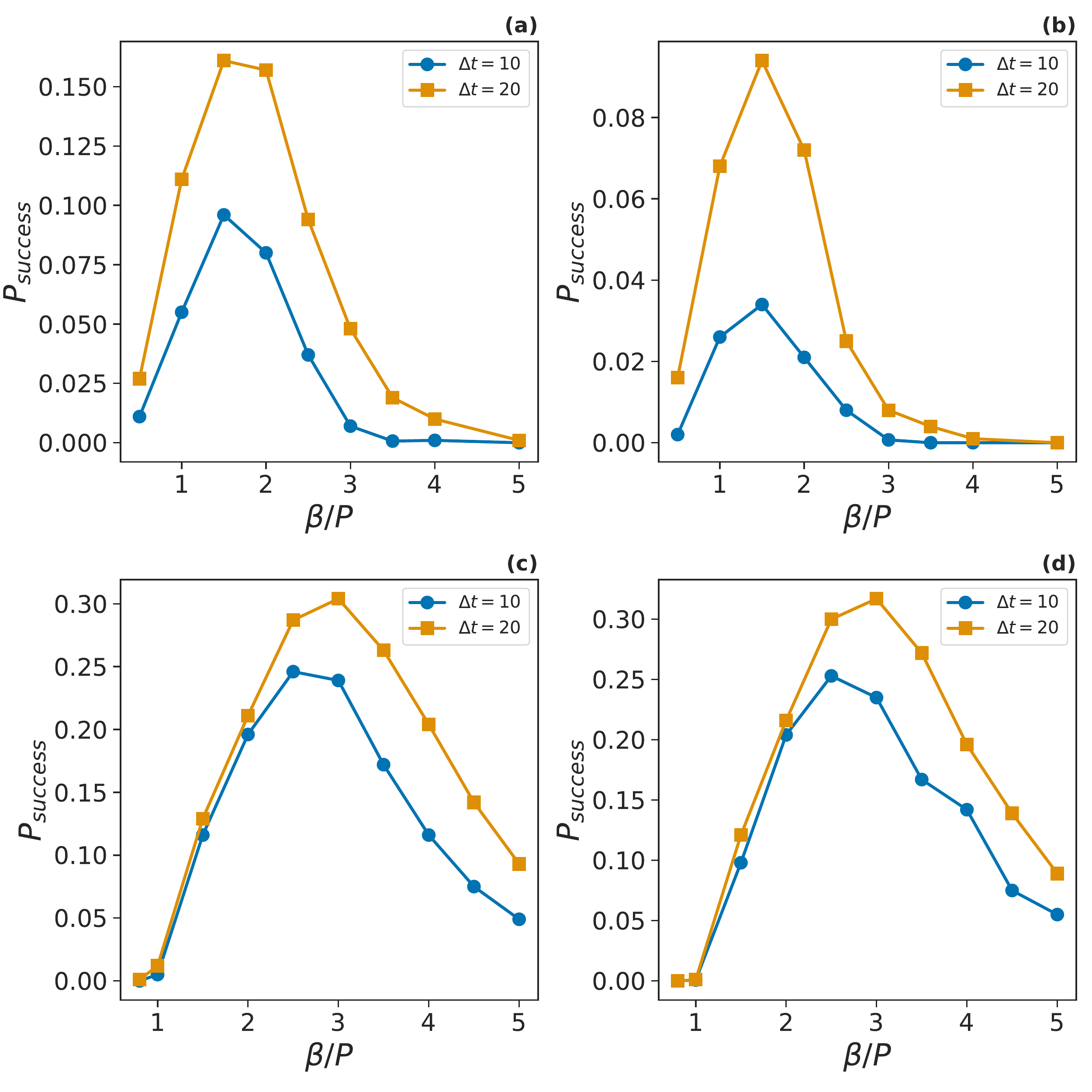} 
\caption{Success probability of the quantum annealing algorithms in the binary perceptron problem. The annealing process is simulated by the quantum Monte Carlo algorithm for the standard ((a),(b)) and reinforced ((c),(d)) QA algorithms. The left and right panels are for $\Gamma=1$ and $\Gamma=2$, respectively. Each data point gives the fraction of successful runs $P_{success}$ in $500-1000$ random instances of the problem with $N=200$ and $M=0.8 N$. The parameters here are the number of replicas in the QMC algorithm $P=100$, the number of time steps $T=100$, the inverse temperature $\beta$, the equilibration time $\Delta t=10,20$, the rate of increasing the reinforcement parameter $\delta r=0.005$, and parameters of the generalized entropy $(q=-4,\lambda=0.5)$}\label{f12}
\end{figure}

Figure \ref{f10} shows the success probability of the above algorithm in the binary perceptron problem with $N=200$ and $M=0.8 N$. In Fig. \ref{f11}, we see how the average minimum energy changes with time step $t$.
For comparison, we also report the minimum energy obtained by the classical simulated annealing (SA) and the standard quantum annealing algorithm.  Here, the set of problem instances and the number of spin flips are the same for all the algorithms \cite{SM}. The SA algorithm starts with inverse temperature $\beta_0=0.01$ which increases linearly with $\delta\beta=10^{-3}$ in $T=10^4$ steps of $\Delta t=10$ Monte Carlo sweeps. This algorithm does not find a solution in $10^3$ random realizations of the perceptron problem. The number of spin flips in the SA algorithm is $T\Delta t N$. Even for $\delta\beta=10^{-4}$ and $T=10^5$ the probability of finding a solution is about $0.01$. On the other hand, the success probability of the reinforced quantum annealing algorithm is about $0.2$, which is nearly $2$ times greater than that of the values we obtain by the standard quantum annealing algorithm. This is observed also in Fig. \ref{f12} which compares the performance of the two algorithms for different values of $\beta$. Table \ref{tab1} shows how the results change with the other parameters of the algorithm.      

\begin{table}[ht]
\begin{tabular}{|l|l|l|l||l|l|}
\hline
$(\lambda, \delta r)$ &  $(0.5, 0.005)$ &  $(0.5, 0.003)$ &  $(0.5, 0.007)$ &  $(0.3, 0.005)$ &  $(0.7, 0.005)$  \\
\hline
$\Gamma=1$ &  $0.19$ &  $0.14$ &  $0.19$ &  $0.08$ &  $0.19$   \\
\hline
$\Gamma=2$ &  $0.20$ &  $0.12$ &  $0.18$ &  $0.06$ &  $0.19$  \\
\hline
\end{tabular}
\caption{Variation of the success probability with the parameters of the rQA algorithm in the perceptron problem. Here $N=200, M=0.8N$ and the results are obtained with the quantum Monte Carlo simulations for $10^3$ random problem instances. The other parameters are $T=100, \Delta t=10, P=100, \beta=200$, and $q=-4$.}\label{tab1}
\end{table}

\section{Conclusion}\label{S2}
We studied a reinforced quantum annealing algorithm in which an entropy function is added to the energy of the classical problem as a reinforcement. The reinforcement is to increase the minimum energy gap in the annealing process and thus enhance the success probability of the algorithm in finding the ground state of the original problem.   
We observed a transition in the optimal parameters of the algorithm as the number of solutions in the quantum search problem was varied. We also observe good performance of the success probability of the algorithm compared to the standard quantum annealing algorithm. Moreover, the reinforcement term is able to change the discontinuous phase transition of the mean-field p-spin model in the quantum annealing algorithm to a continuous transition. That is a qualitative change in the scaling of the minimum energy gap (from exponential to polynomial in system size) which controls the efficiency of the quantum annealing algorithm in the absence of degeneracy. Finally, we employed a quantum Monte Carlo algorithm to simulate the annealing process in the binary perceptron problem. We found that reinforcement can improve on the success probability of the standard quantum annealing algorithm close to the threshold capacity of the perceptron.

The main point here is that the entropy is a nonlocal function and depends on the physical state of the system in the annealing process. This in particular makes the practical implementation of the exact algorithm very challenging. To ease the first problem, one may use an approximate expression for the reinforced Hamiltonian considering only one- and two-spin effective interactions. More specifically, this can be done in the limit $Q\to 0$ when high order correlations in the wave function are ignored. On the other hand, many identical copies of the system are needed to infer the quantum state of the system in a sequence of weak quantum measurements. To get around this problem, in Sec. \ref{S11} we showed that the minimum energy gap can also be increased when we replace the quantum expectations with the equivalent thermal ones. The latter can even be estimated in a local way within the Bethe approximation as described in Appendix \ref{app1}. Moreover, as we saw in the perceptron problem, even a mean-field approximation of the entropy function in the reinforced quantum annealing algorithm displays good performance compared to the standard algorithm. It would be interesting to see how more accurate expressions for the local entropy work in other computationally hard optimization problems.

\acknowledgments
I would like to thank M. H. Zarei for helpful discussions. This work was performed using the ALICE compute resources provided by Leiden University.

\appendix

\section{The free energy approximation}\label{app1}
We can rewrite the entropy function as
\begin{align}
S_q(\boldsymbol\sigma:\lambda)=-\frac{1}{Q}\left(e^{-Q\beta[F(\boldsymbol\sigma:\lambda)-F(0)]}-1\right),
\end{align}
where
\begin{align}
e^{-\beta F(\boldsymbol\sigma:\lambda)}=\sum_{\boldsymbol\sigma'}e^{-\beta E(\boldsymbol\sigma')-\lambda D(\boldsymbol\sigma',\boldsymbol\sigma)}.
\end{align}
Now for $Q=q/N$:
\begin{align}
s_q(\boldsymbol\sigma:\lambda)=\frac{S_q(\boldsymbol\sigma:\lambda)}{N}=-\frac{1}{q}\left(e^{-q\beta[f(\boldsymbol\sigma:\lambda)-f(0)]}-1\right).
\end{align}

For a local energy function $E(\boldsymbol\sigma)=\sum_a e_a(\sigma_{\partial a})$ one can use the Bethe approximation to find an estimation of the free energy \cite{MM-book-2009},
\begin{align}
F(\boldsymbol\sigma:\lambda)=\sum_a \Delta F_a+\sum_i \Delta F_i-\sum_{(ai)} \Delta F_{ai},
\end{align}
where the local free energies are given by
\begin{align}
e^{-\beta\Delta F_a} &=\sum_{\sigma_{\partial a}'} e^{-\beta e_a(\sigma_{\partial a}')}\prod_{i\in \partial a}\mu_{i\to a}(\sigma_i'),\\
e^{-\beta\Delta F_i} &=\sum_{\sigma_i'} e^{-\lambda (\sigma_i'-\sigma_i)^2}\prod_{a\in \partial i}\mu_{a\to i}(\sigma_i'),\\
e^{-\beta\Delta F_{ai}} &=\sum_{\sigma_i'} \mu_{a\to i}(\sigma_i')\mu_{i\to a}(\sigma_i').
\end{align}
Note that $f(0)=F(\boldsymbol\sigma:0)/N$ and $f(\boldsymbol\sigma:\lambda)=F(\boldsymbol\sigma:\lambda)/N$.
The cavity messages are obtained by solving the Belief Propagation equations, 
\begin{align}
\mu_{a\to i}(\sigma_i') &\propto \sum_{\sigma_{\partial a\setminus i}'} e^{-\beta e_a(\sigma_{\partial a}')}\prod_{j\in \partial a\setminus i}\mu_{j\to a}(\sigma_j'),\\
\mu_{i\to a}(\sigma_i') &\propto  e^{-\lambda  (\sigma_i'-\sigma_i)^2}\prod_{b\in \partial i\setminus a}\mu_{b\to i}(\sigma_i').
\end{align}
In this way, one obtains an approximate free energy which is expected to be asymptotically exact for locally tree-like interaction graphs.

\section{A local representation of the perceptron problem}\label{app2}
Consider a chain of $\sigma_i$ variables from $i=1,\dots,N$ which is replicated $M$ times as depicted in Fig. \ref{f9}. The $\sigma_i^a$ variables are assumed to have the same values in all replicas. This can be achieved by application of strong ferromagnetic interactions between the adjacent replicas. Let us define
a set of left-to-right messages $m_{i}^a=m_{i-1}^a+\sigma_i\xi_i^a$ for all $i=1,\dots,N-1$ and $a=1,\dots,M$ with $m_{0}^a=0$. Similarly, we define a set of right-to-left messages $n_{i}^a=n_{i+1}^a+\sigma_i\xi_i^a$ for all $i=2,\dots,N-1$ and $a=1,\dots,M$ with $n_{N}^a=0$. Note that the $m_i^a$ and $n_i^a$ take values in $(-N,\dots,+N)$. Then the energy function reads
\begin{align}
E(\vec{\boldsymbol\sigma},\vec{\mathbf{m}},\vec{\mathbf{n}}:\vec{\boldsymbol\xi},\mathbf{s})=\frac{1}{N}\sum_a\sum_i e(\sigma_i^a,m_{i-1}^a,n_{i+1}^a:\xi_i^a,s^a),
\end{align}
where the local energy is
\begin{align}
e(\sigma_i^a,m_{i-1}^a,n_{i+1}^a:\xi_i^a,s^a)=1-s^a\mathrm{sgn}(m_{i-1}^a+n_{i+1}^a+\sigma_i\xi_i^a).
\end{align}
That is the total energy is written as a local function of the weights $\sigma_i^a$ and the auxiliary variables $(m_i^a, n_i^a)$.

\end{document}